\documentclass[aip,apl,reprint]{revtex4-1}

\usepackage{graphicx}

\begin{document}

\preprint{}

\title{A Compact, Transportable, Microchip-Based System for High Repetition Rate Production of Bose-Einstein Condensates}


\author{Daniel M. Farkas}
\thanks{Author to whom correspondence should be addressed. Electronic mail: daniel.farkas@colorado.edu.}
\author{Kai M. Hudek}
\author{Evan A. Salim}
\author{Stephen R. Segal}
\author{Matthew B. Squires}
\thanks{Current affiliation: Air Force Research Laboratory, Hanscom AFB, Massachusetts 01731, USA.}
\author{Dana Z. Anderson}
\affiliation{JILA and Department of Physics, University of Colorado, Boulder, CO 80309}

\date{\today}

%%%%%%%%%%%%%%%%%%%%%%%%%%%%%%%%%%%%%%%%%%%%%%%%%%%%%%%%%%%%%%%%%%%%%%%%%%%%%%%%%%%%%%%%%%%%%%%%%%%%%%%%%%%%%%%%%%%%%%%%%%%%%%%%%%%%%%%%%%%%%%%%%%%%%%

\begin{abstract}
We present a compact, transportable system that produces Bose-Einstein condensates (BECs) near the surface of an integrated atom microchip. The system occupies a volume of 0.4\,m$^3$ and operates at a repetition rate as high as 0.3\,Hz. Evaporative cooling in a chip trap with trap frequencies of several kHz leads to nearly pure condensates containing 1.9$\times 10^4$ $^{87}$Rb atoms. Partial condensates are observed at a temperature of 1.58(8)\,$\mu$K, close to the theoretical transition temperature of 1.1\,$\mu$K.
\end{abstract}

%\keywords{}

\maketitle

%%%%%%%%%%%%%%%%%%%%%%%%%%%%%%%%%%%%%%%%%%%%%%%%%%%%%%%%%%%%%%%%%%%%%%%%%%%%%%%%%%%%%%%%%%%%%%%%%%%%%%%%%%%%%%%%%%%%%%%%%%%%%%%%%%%%%%%%%%%%%%%%%%%%%%

Since the first experimental demonstrations of Bose-Einstein condensation (BEC) in a gas of neutral atoms,~\cite{anderson1995, davis_bose-einstein_1995, bradley_bose-einstein_1997} studies of BEC and related forms of ultracold matter have been largely motivated by purely scientific interests. The complexity and size of the required apparatus necessitate that these experiments remain confined to research laboratories. However it has become increasingly evident that ultracold matter can play a utilitarian role in applications such as atomic clocks, inertial sensors, and electric and magnetic field sensing.~\cite{treutlein_coherence_2004, mcguirk_sensitive_2002, yu_development_2006, Konemann2007, stern_light-pulse_2009, Wildermuth2006} Indeed, much of the work on ultracold atom chip technology is predicated on the need for compact systems that can find their way out of the laboratory and into the field.

\begin{figure}
\includegraphics[width=3.3in]{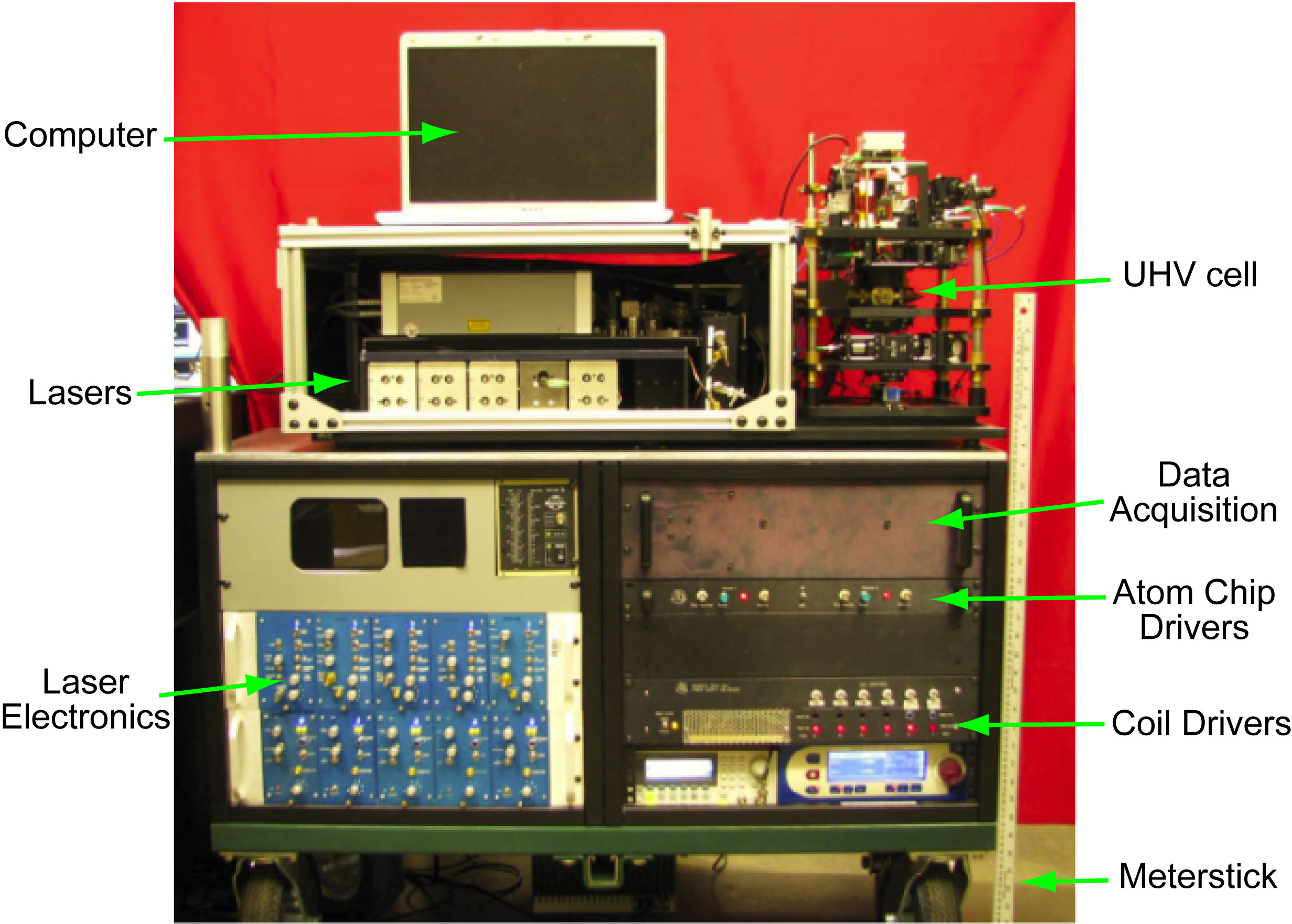}
\caption{\label{CartPicture}(color online) Picture of a compact, transportable system that produces and images BECs.}
\end{figure}

We present here a compact, movable, microchip-based BEC production system that occupies a volume of 0.4\,m$^3$, operates at a repetition rate as high as 0.3\,Hz, and produces BECs containing 1.9$\times10^4$ atoms in the $|F=2,m_F=2\rangle$ ground state of $^{87}$Rb (see Fig.~\ref{CartPicture}). The system contains \emph{all} of the components needed to produce and image BECs, including an ultra-high vacuum (UHV) system, lasers, data acquisition hardware, electronics, and imaging equipment. The system can be easily reconfigured for use with atom chips having unique wire patterns designed for different applications. As such, it can serve as a standardized platform for a variety of portable experiments that utilize ultracold matter.

\begin{figure}
\includegraphics[width=2in]{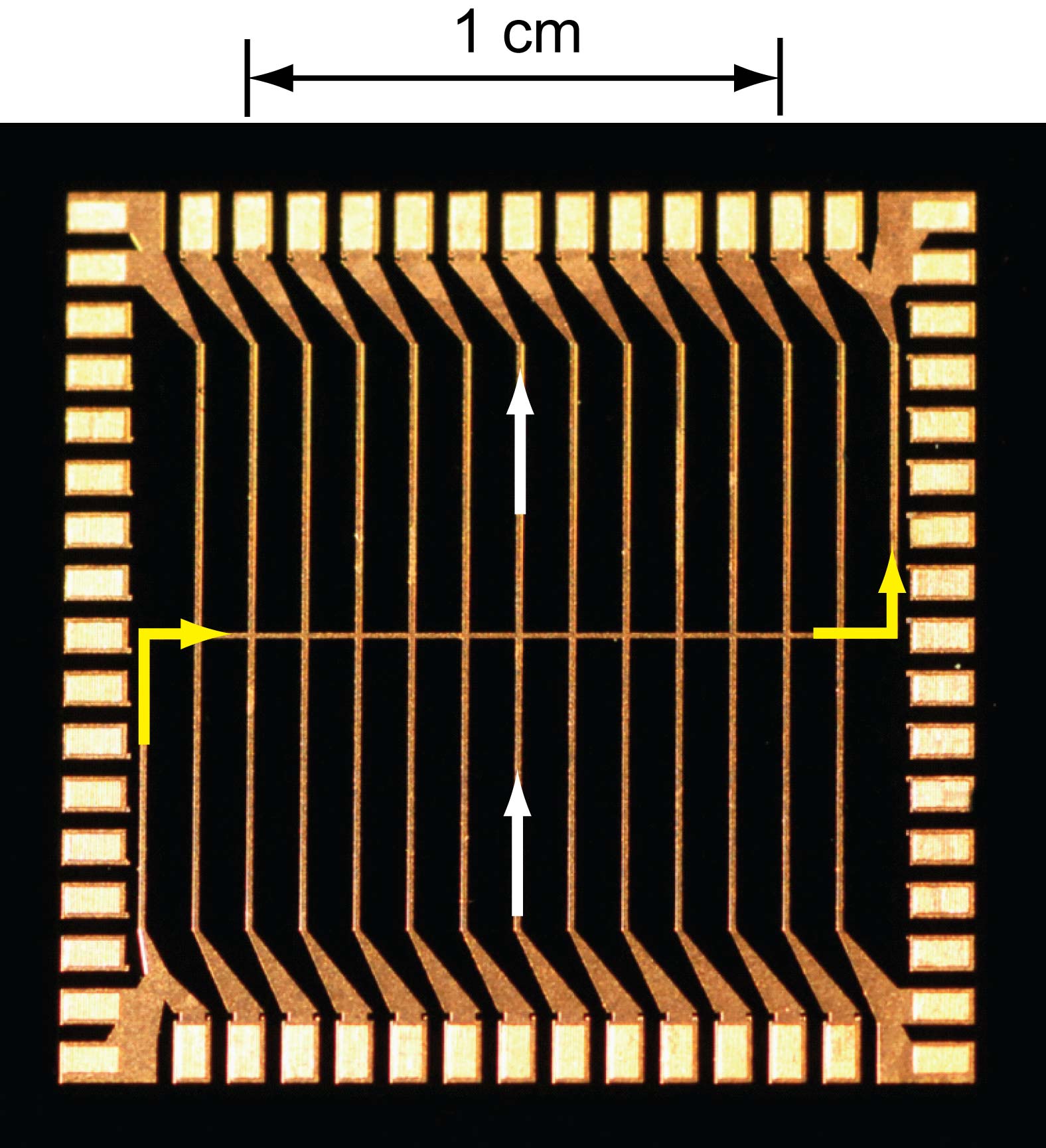}
\caption{\label{AtomChip}(color online) Picture of the vacuum side of the atom microchip. Arrows represent the flow of current through the Z wire (yellow) and dimple wire (white).}
\end{figure}

Significant reductions in power consumption and volume are achieved by trapping atoms with a microchip rather than ``traditional," macroscopic-sized magnetic coils.\cite{hansel_bose-einstein_2001, ott_bose-einstein_2001,Horikoshi2006,du_atom-chip_2004} Patterned using standard fabrication techniques by Teledyne Scientific and Imaging, LLC, the 26\,mm$\times$26\,mm atom chip used here is formed by depositing 100\,$\mu$m-wide, 10\,$\mu$m-thick copper conductors onto a 450\,$\mu$m-thick silicon substrate (see Fig.~\ref{AtomChip}). A segment in a ``Z" configuration (called the Z-wire) is used with external bias fields to create a Ioffe-Pritchard trap.\cite{reichel_microchip_2002} Higher trap frequencies are obtained with a dimple trap that is created by running an additional current perpendicular to the center of the Z-wire. The chip is anodically bonded to the UHV cell, where it functions as a structural wall of the vacuum system.\cite{du_atom-chip_2004} Atoms are trapped at distances less than 200\,$\mu$m from the room-temperature surface of the chip and less than 1\,mm away from the ambient environment. Current is passed into the vacuum system through hermetic, UHV-compatible, through-chip vias that serve as electrical feedthroughs. Each via can sustain 2.5\,A of current for several hundred milliseconds without destructively overheating. The chip conductors are driven by current servos that are powered by floating, switched-mode power supplies.

%%%%%%%%%%%%%%%%%%%%%%%%%%%%%%%%%%%%%%%%%%%%%%%%%%%%%%%%%%%%%%%%%%%%%%%%%%%%%%%%%%%%%%%%%%%%%%%%%%%%%%%%%%%%%%%%%%%%%%%%%%%%%%%%%%%%%%%%%%%%%%%%%%%%%%

\begin{figure}
\includegraphics[width=3in]{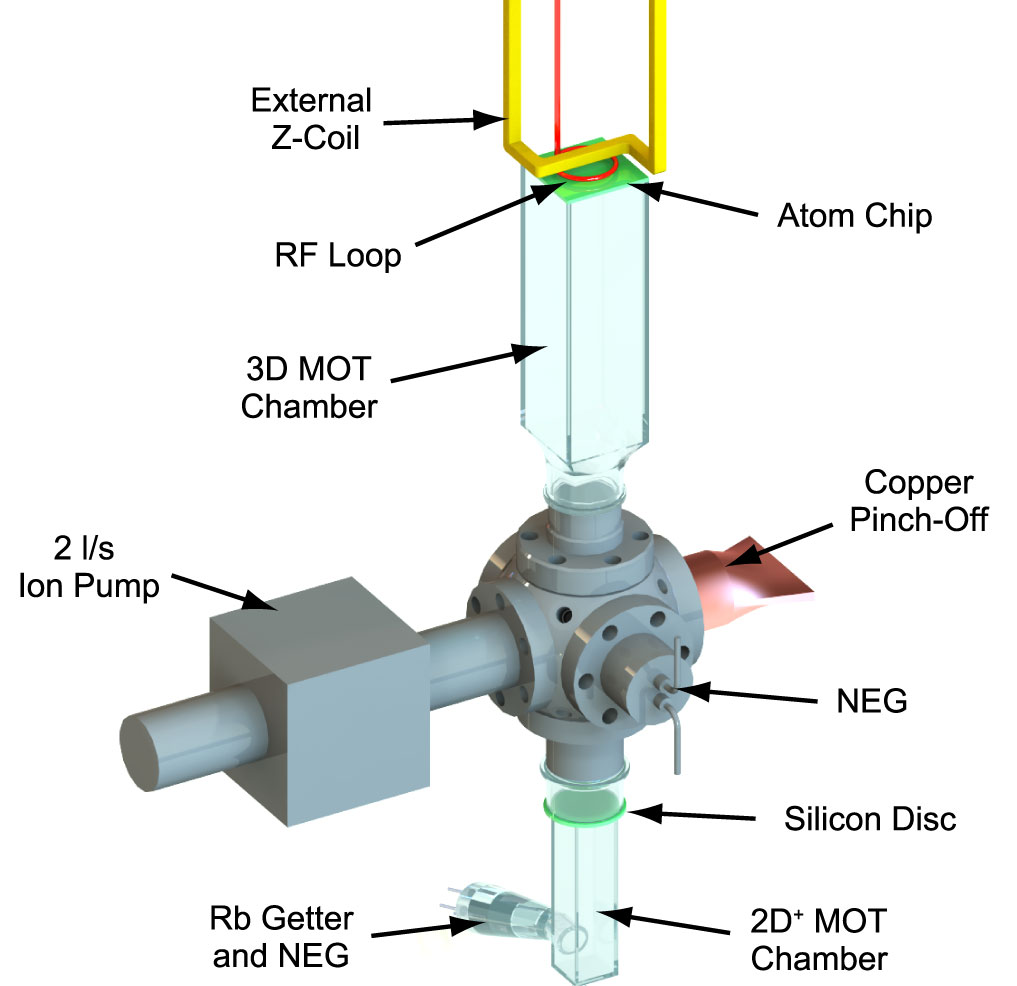}
\caption{\label{TwoChamber}(color online) Schematic of the two-chamber UHV vacuum system, RF loop, and external Z-coil. The distance from the bottom of the 2D$^+$ MOT chamber to the top of the atom chip is 27\,cm.}
\end{figure}

To achieve high repetition rates, atoms must be quickly loaded into a trap from a high-pressure background vapor. However, to minimize the loss of trapped atoms from background collisions, a low pressure is also required. A standard solution to meeting these disparate requirements is a two-chamber system: one chamber contains a high-pressure background vapor of the atomic species of interest while a second chamber is pumped to maintain a low background pressure. Our implementation uses chambers formed from glass fluorimeter cells (see Fig.~\ref{TwoChamber}). In the bottom chamber, a getter source creates a Rb pressure between $10^{-8}$ and $10^{-7}$\,Torr. Differential pumping between this chamber and the rest of the system is achieved with a 750\,$\mu$m hole drilled into a silicon disc that is anodically bonded to the lower chamber. UHV pressures are maintained in the upper chamber with a 2\,l/s ion pump and a non-evaporable getter (NEG).

In the lower chamber, atoms are loaded into a two-dimensional (2D) magneto-optical trap (MOT) using the quadrupole field created by permanent magnets. The MOT is unconfined in the vertical direction, and cold Rb atoms are pushed upward through the hole in the silicon disc by a cooling laser beam. Reflection of this laser beam off the silicon disc around the hole provides additional cooling in the lower chamber, thereby forming a 2D$^+$ MOT.~\cite{dieckmann_two-dimensional_1998} In the upper chamber, cold atoms are loaded into a six-beam, three-dimensional (3D) MOT using the magnetic quadrupole field generated by a pair of anti-Helmholtz coils. Loading rates into the 3D MOT of $10^{9}$ atom/s have been achieved while the number of atoms in the 3D MOT typically saturates at 2$\times10^{9}$ atoms.

Five distributed-feedback laser diodes at 780\,nm are used for cooling, repumping, optical pumping, and imaging. Two of the lasers are locked to spectral lines in Rb vapor cells, and serve as masters for the cooling and repump transitions. Cooling and repump slave lasers are offset-locked from their respective master lasers by stabilizing the RF heterodyne beat between each master and slave. Voltage-controlled oscillators set the offset frequencies. A combined 7\,mW of light from the cooling and repump slave lasers (85\% cooling, 15\% repump) is coupled into a tapered amplifier (Sacher TEC-400). This output is split into two, yielding 70\,mW and 50\,mW of light for the 2D$^+$ MOT and 3D MOT, respectively. The fifth laser is offset-locked to the cooling master and used for optical pumping and absorption imaging.

%%%%%%%%%%%%%%%%%%%%%%%%%%%%%%%%%%%%%%%%%%%%%%%%%%%%%%%%%%%%%%%%%%%%%%%%%%%%%%%%%%%%%%%%%%%%%%%%%%%%%%%%%%%%%%%%%%%%%%%%%%%%%%%%%%%%%%%%%%%%%%%%%%%%%%

While pre-cooling atoms in the 2D$^+$ MOT, the 3D MOT is filled for 1 to 2\,s to 5$\times10^8$ atoms. The atomic cloud is then spatially compressed by increasing the cooling slave laser detuning from $-2.5\Gamma$ to $-3.5\Gamma$~(where $\Gamma$=6.0\,MHz is the natural linewidth of the cooling transition), blue-detuning the repump slave laser by 200\,MHz, and increasing the magnetic field gradient from 10\,G/cm to 30\,G/cm. The atoms are further cooled with 4 to 7\,ms of sub-Doppler polarization gradient cooling to temperatures below 20\,$\mu$K. Circularly polarized light optically pumps the atoms into the $|F=2,m_F=2\rangle$ ground state.

The atoms are then transported vertically and loaded into the chip trap using an external Z-coil placed directly above the atom chip (see Fig.~\ref{TwoChamber}). The coil's windings form a ``Z" configuration in the same sense as the ``Z" pattern on the atom chip. In conjunction with external bias fields, the Z-coil creates a Ioffe-Pritchard trap that is well mode-matched to the chip-Z trap, permitting efficient, adiabatic transfer. The atoms are initially caught 1.5\,cm below the atom chip, and are then moved upward adiabatically by reducing the coil current. The atoms are transferred onto the chip by ramping off the external Z-coil while ramping up the chip-Z current to 5\,A. The atoms are then loaded into the dimple trap by reducing the chip-Z current to 3.25\,A while increasing the dimple current to 1.25\,A. The resulting dimple trap is centered 115\,$\mu$m below the chip surface, has calculated trap frequencies of 6.7\,kHz$\times$6.7\,kHz$\times$610\,Hz, and contains 3$\times 10^7$ atoms. A dimple trap lifetime of 6.5\,s was measured.

\begin{figure*}
\includegraphics[width=7in]{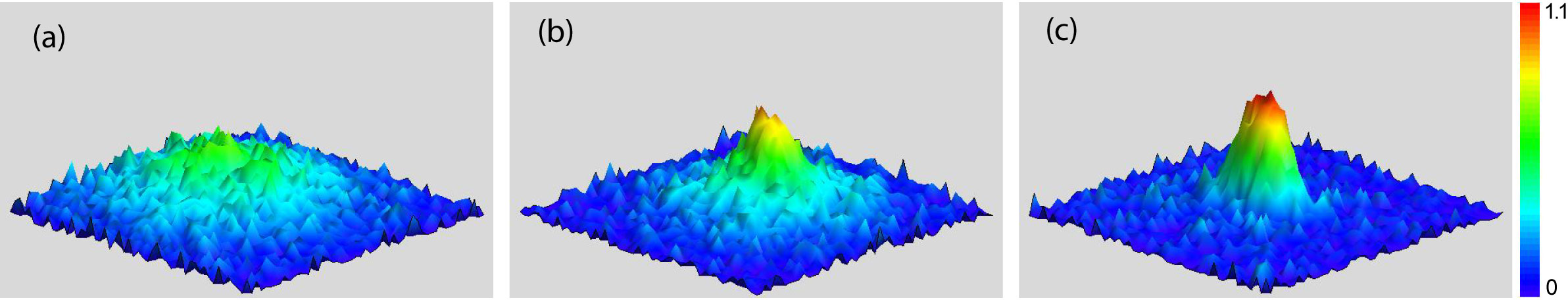}
\caption{\label{BECformation}(color online) OD distributions demonstrating the onset of condensation: (a) a cloud of 6.9$\times 10^4$ non-condensed atoms at a temperature of 1.92(6)\,$\mu$K; (b) a partially condensed cloud of 5$\times 10^4$ atoms at 1.58(8)\,$\mu$K; (c) a nearly pure condensate of 1.9$\times 10^4$ atoms. Images were taken after 5\,ms of free expansion.}
\end{figure*}

Several stages of RF evaporative cooling reduce the cloud temperature below the transition point for BEC formation. The RF field is produced by driving a small loop mounted onto the end of the external-Z coil frame (see Fig.~\ref{TwoChamber}). The RF frequency is swept linearly in five stages, totaling 1.3 to 2.5\,s in duration, with each stage cutting halfway into the remaining trap. To avoid excessive three-body losses, the trap is reduced after the first stage by adiabatically decreasing the magnetic bias fields; the reduced trap has calculated frequencies of 2.4\,kHz$\times$2.3\,kHz$\times$340\,Hz and the trap center is shifted to 170\,$\mu$m below the chip surface.

After evaporation, the atomic cloud is prepared for imaging by again reducing the trap. This final trap has calculated frequencies of 1.2\,kHz$\times$1.2\,kHz$\times$200\,Hz and is centered 260\,$\mu$m below the chip surface. The cloud is then dropped by turning off all magnetic fields. After a variable time-of-flight, a picture of the cloud is obtained via absorption imaging on a CCD camera (Basler A102f). The images are used to calculate the 2D optical depth (OD), which is proportional to the integrated column density of the atomic cloud. The data is fit to one of three functions, depending on whether or not condensation has been achieved. In the absence of condensation, the cloud consists entirely of thermal atoms that form a Bose-enhanced Gaussian distribution. For complete condensates, assuming the Thomas-Fermi limit, the cloud has the shape of an inverted paraboloid. For clouds containing both thermal and condensed atoms, the data is fit to the sum of these two functions.

%%%%%%%%%%%%%%%%%%%%%%%%%%%%%%%%%%%%%%%%%%%%%%%%%%%%%%%%%%%%%%%%%%%%%%%%%%%%%%%%%%%%%%%%%%%%%%%%%%%%%%%%%%%%%%%%%%%%%%%%%%%%%%%%%%%%%%%%%%%%%%%%%%%%%%

The onset of condensation can be seen in the OD distributions of Fig.~\ref{BECformation}. In (a), the RF evaporative sweep stops at 90\,kHz above the trap bottom, resulting in a cloud of 6.9$\times 10^4$ non-condensed atoms at a temperature of 1.92(6)\,$\mu$K. The beginning of condensation is shown in (b), where the RF sweep stops at 60\,kHz above the trap bottom. The cloud of 5$\times 10^4$ atoms has a temperature of 1.58(8)\,$\mu$K, slightly above than the calculated transition temperature of 1.1\,$\mu$K. Here, the condensate is evident by the higher peak OD and smaller Gaussian wings. In (c), where the RF sweep stops at 30\,kHz above the trap bottom, the lack of Gaussian wings indicates a nearly pure condensate of 1.9$\times 10^4$ atoms.

%%%%%%%%%%%%%%%%%%%%%%%%%%%%%%%%%%%%%%%%%%%%%%%%%%%%%%%%%%%%%%%%%%%%%%%%%%%%%%%%%%%%%%%%%%%%%%%%%%%%%%%%%%%%%%%%%%%%%%%%%%%%%%%%%%%%%%%%%%%%%%%%%%%%%%

In conclusion, we have used a compact, transportable, microchip-based system to produce BECs containing 1.9$\times 10^4$ $^{87}$Rb atoms. Containing all of the components needed to produce and image BECs, the system occupies a total volume of 0.4\,m$^3$ and operates at a repetition rate of 0.3\,Hz. The system is an ideal platform for utilizing ultracold matter for portable applications.

We are grateful to L. Czaia for her capable assistance with vacuum cell fabrication and to F. Majdeteimouri for his substantial contributions to software development for the control system. We are grateful to J. DeNatale and R. Mihaliovich of Teledyne Scientific and Imaging, LLC, for their contributions to the development of atom chip technologies. We are grateful to M. Anderson and B. Luey of Vescent Photonics, Inc., for their knowledgeable assistance with the laser systems. Finally, we are grateful to S. McBride, J. Michalchuk and D. Ackerman of Sarnoff Corporation for useful discussions and their contributions to the laser system integration. This work was supported in part by the Defense Advanced Research Projects Agency, the Army Research Office (W911NF-04-1-0043), and the National Science Foundation through a Physics Frontier Center (PHY0551010).

\end{document}